\documentclass[twocolumn]{webofc}

\usepackage[varg]{txfonts}   
\usepackage{bm}
\begin{document}
\title{Evolution of nuclear spin-orbit splittings with Skyrme functional SAMi-T}
%
%
\author{\firstname{Shihang} \lastname{Shen}\inst{1,2}\fnsep \and
        \firstname{Gianluca} \lastname{Col\`o}\inst{1,2}\fnsep\thanks{\email{Gianluca.Colo@mi.infn.it}} \and
        \firstname{Xavier} \lastname{Roca-Maza}\inst{1,2}\fnsep
}

\institute{Dipartimento di Fisica, Universit\`a degli Studi di Milano, Via Celoria 16, I-20133 Milano, Italy
\and
           INFN, Sezione di Milano, Via Celoria 16, I-20133 Milano, Italy
          }

\abstract{%
  A new Skyrme functional has been developed with tensor term guided by \textit{ab initio} relativistic Brueckner-Hartree-Fock (RBHF) studies on neutron-proton drops.
  Instead of extracting information on the tensor force from experimental single-particle energy splittings, the RBHF calculations do not contain beyond mean-field effects such as particle-vibration coupling and therefore the information on the tensor force can be obtained without ambiguities.
  The new functional gives a good description of nuclear ground-state properties as well as various giant resonances.
  The description for the evolution of single-particle energy splittings is also improved by the new functional.
}
\maketitle
\section{Introduction}\label{intro}

Nuclear density functional theory is a powerful tool to describe various nuclear phenomena along a large part of the nuclear chart \cite{Bender2003,Meng2016}.
However, there are still many open questions concerning current nuclear energy density functionals such as the tensor force term \cite{Sagawa2014}.
When constructing the functional by fitting its parameters against bulk properties such as binding energies and charge radii, tensor force plays relatively small role and is difficult to be determined.
Even though it shows important effects in describing the evolution of single-particle levels along isotopic (or isotonic) chains \cite{Schiffer2004,Otsuka2005,Brown2006,Colo2007,Brink2007,Lesinski2007,LongWH2008}, it is very much mixed with the beyond-mean-field effect, such as particle-vibration coupling \cite{Afanasjev2015}.

Taking advantage of the advances of nuclear \textit{ab initio} calculations, a clear sign of tensor force has been shown in the evolution of spin-orbit (s.o.) splittings in the neutron drops \cite{Shen2018,Shen2018a}.
The adopted \textit{ab initio} framework, the relativistic Brueckner-Hartree-Fock theory \cite{Shen2016,Shen2017,Shen2018b,Shen2019a}, has no beyond-mean-field effects such as particle-vibration coupling and therefore the information of tensor force has no ambiguities.
It provides a important guide to constrain the tensor force in the nuclear energy density functional.
Along this direction, in order to extract also the information of neutron-proton tensor force we have extended the study to neutron-proton drops, which is also an ideal system confined in an external field without consideration of center-of-mass correction nor Coulomb interaction \cite{Shen2019}.
Then we developed a new Skyrme functional, SAMi-T, with tensor term by fitting to the evolution of s.o. splittings of the neutron-proton drops calculated by RBHF \cite{Shen2019}.

In Sec.~\ref{sec-1} we briefly give the formalism of the tensor term in Skyrme functional.
Results of evolution of s.o. splittings along the $Z = 50$ isotopic chain and $N = 82$ isotonic chain are given in Sec.~\ref{sec-2}.
Finally, we give the summary in Sec.~\ref{sec-3}.

\section{Formalisms}\label{sec-1}
The Skyrme effective interaction with two-body tensor term $V_T$ is written in the standard form as~\cite{Vautherin1972,Stancu1977}.
\begin{align}
V(\mathbf{r}_1,\mathbf{r}_2) &= t_0(1+x_0P_\sigma) \delta(\mathbf{r}) + \frac{1}{2}t_1
(1+x_1P_\sigma) \left[ {\mathbf{P}'}^2\delta(\mathbf{r}) + \delta(\mathbf{r}) \mathbf{P}^2 \right] \notag \\
&~~~ + t_2(1+x_2P_\sigma) \mathbf{P}' \cdot \delta(\mathbf{r}) \mathbf{P} + \frac{1}{6}t_3 (1+x_3P_\sigma) \rho^\gamma(\mathbf{R}) \delta(\mathbf{r}) \notag \\
&~~~ + iW_0(\bm{\sigma}_1+\bm{\sigma}_2) \cdot \left[ \mathbf{P}'\times \delta(\mathbf{r}) \mathbf{P} \right] + V_T(\mathbf{r}_1,\mathbf{r}_2), \label{eq:vskyrme} \\
V_T(\mathbf{r}_1,\mathbf{r}_2) &= \frac{T}{2} \left\{\left[(\bm{\sigma}_1\cdot\mathbf{P}') (\bm{\sigma}_2\cdot\mathbf{P}')-\frac{1}{3}(\bm{\sigma}_1\cdot\bm{\sigma}_2) {\mathbf{P}'}^{2}\right]\delta(\mathbf{r}) \right. \notag \\
&~~~ \left. +\delta(\mathbf{r})\left[(\bm{\sigma}_1\cdot\mathbf{P})
(\bm{\sigma}_2\cdot\mathbf{P})-\frac{1}{3}(\bm{\sigma}_1\cdot\bm{\sigma}_2)
{\mathbf{P}}^{2}\right]\right\} \notag \\
&~~~ +U\left\{(\bm{\sigma}_1\cdot\mathbf{P}')\delta(\mathbf{r})
(\bm{\sigma}_2\cdot\mathbf{P})-\frac{1}{3}(\bm{\sigma}_1\cdot\bm{\sigma}_2)
\left[\mathbf{P}'\cdot\delta(\mathbf{r})\mathbf{P}\right]\right\},
\end{align}
where $\mathbf{r} = \mathbf{r}_1 - \mathbf{r}_2, \mathbf{R} = \frac{1}{2}(\mathbf{r}_1 + \mathbf{r}_2), \mathbf{P} = \frac{1}{2i}(\nabla_1-\nabla_2)$, and $\mathbf{P}'$ is the hermitian conjugate of $\mathbf{P}$ acting on the left.
The spin-exchange operator reads $P_\sigma = \frac{1}{2}(1+\bm{\sigma}_1\cdot\bm{\sigma}_2)$, and $\rho$ is the total nucleon density.

The Hartree-Fock equations for each s.p. level can be obtained by the variational method with respect to the HF total energy as
\begin{equation}\label{eq:}
  \left[ -\frac{\hbar^2}{2M}\nabla^2 + U_q(\mathbf{r}) \right] \psi_k(\mathbf{r}) = e_k \psi_k(\mathbf{r}),
\end{equation}
where $e_k$ is the single-particle energy, $\psi_k$ is the corresponding wave function, 
and $q = 0(1)$ labels neutrons (protons).
The single-particle potential $U_q(\mathbf{r})$ is a sum of central, Coulomb and spin-orbit terms,
\begin{equation}\label{eq:}
  U_q(\mathbf{r}) = U_{q}^{\rm(c)}(\mathbf{r}) + \delta_{q,1}U_{C}(\mathbf{r}) + \mathbf{U}_{q}^{\rm(s.o.)}(\mathbf{r}) \cdot (-i)(\nabla\times\bf{\sigma}).
\end{equation}
The spin-orbit term reads \cite{Stancu1977,Sagawa2014}
\begin{equation}\label{eq:Uso}
  \mathbf{U}_q^{\rm(s.o.)}(\mathbf{r}) = \frac{1}{2} \left[ W_0\nabla\rho + W_0'\nabla\rho_q \right]
  + \left[ \alpha \mathbf{J}_q + \beta\mathbf{J}_{1-q} \right],
\end{equation}
where $\mathbf{J}(\mathbf{r})$ the spin-orbit density. 
Starting from Eq.~(\ref{eq:vskyrme}) one would derive $W_0' = W_0$.
A more general form has been adopted in which $W_0'$ can be defined and fitted independently as in the case of SAMi functional \cite{Roca-Maza2012} among the others.

The parameters $\alpha$ and $\beta$ in Eq.~(\ref{eq:Uso}) include contributions from 
the exchange part of the central term and from the tensor term,
\begin{equation}\label{eq:alpha-beta}
  \alpha = \alpha_c + \alpha_T,\quad \beta = \beta_c + \beta_T,
\end{equation}
where
\begin{subequations}\label{eq:}\begin{align}
  \alpha_c &= \frac{1}{8}(t_1-t_2) - \frac{1}{8}(t_1x_1+t_2x_2),\quad
  \alpha_T = \frac{5}{12}U,\\
  \beta_c &= -\frac{1}{8}(t_1x_1+t_2x_2),\quad
  \beta_T = \frac{5}{24}(T+U).
\end{align}\end{subequations}

For open-shell nuclei, HF + BCS is performed to taken into account the pairing correlations.
The pairing force used in this work is a zero-range, density-dependent one of the form
\begin{equation}\label{eq:}
  V = V_0 \left[ 1- \left( \frac{\rho\left( \frac{\mathbf{r}_1+\mathbf{r}_2}{2} \right)}{\rho_0} \right)^\sigma \right] \delta(\mathbf{r}_1-\mathbf{r}_2),
\end{equation}
with parameters the same as the ones in Ref.~\cite{Colo2007}, $V_0 = 680$ MeV fm$^3$, $\sigma = 1$, and $\rho_0 = 0.16$ fm$^{-1}$.

\section{Results and Discussion}\label{sec-2}

The Skyrme functional SAMi-T has been developed with tensor terms guided by the RBHF calculations of the neutron-proton drops \cite{Shen2019}.
In the RBHF calculation, the Bonn A interaction has been used, which was fitted to the two-nucleon scattering data and to the deuteron properties \cite{Machleidt1989}.
The in-medium interaction ($G$-matrix) is obtained by solving the Bethe-Goldstone equation and then used in the mean-field calculation.
RBHF belongs to the \textit{ab initio} methods as it is a well-controlled many-body approximation in which a realistic interaction in input.
For more details of the RBHF framework used, see Ref.~\cite{Shen2017}.

The ground state properties of nuclei such as binding energies, charge radii, and spin-orbit splittings can be well described.
Furthermore, SAMi-T also keeps the good merit of SAMi \cite{Roca-Maza2012} that it provides a good description also for various collective excitations such as the giant monopole, dipole resonances, and the Gamow-Teller resonances.
Especially for the spin-dipole resonances the description by SAMi-T has been improved with the inclusion of tensor term comparing with SAMi.

\begin{figure}[h]
\centering
\includegraphics[width=7cm]{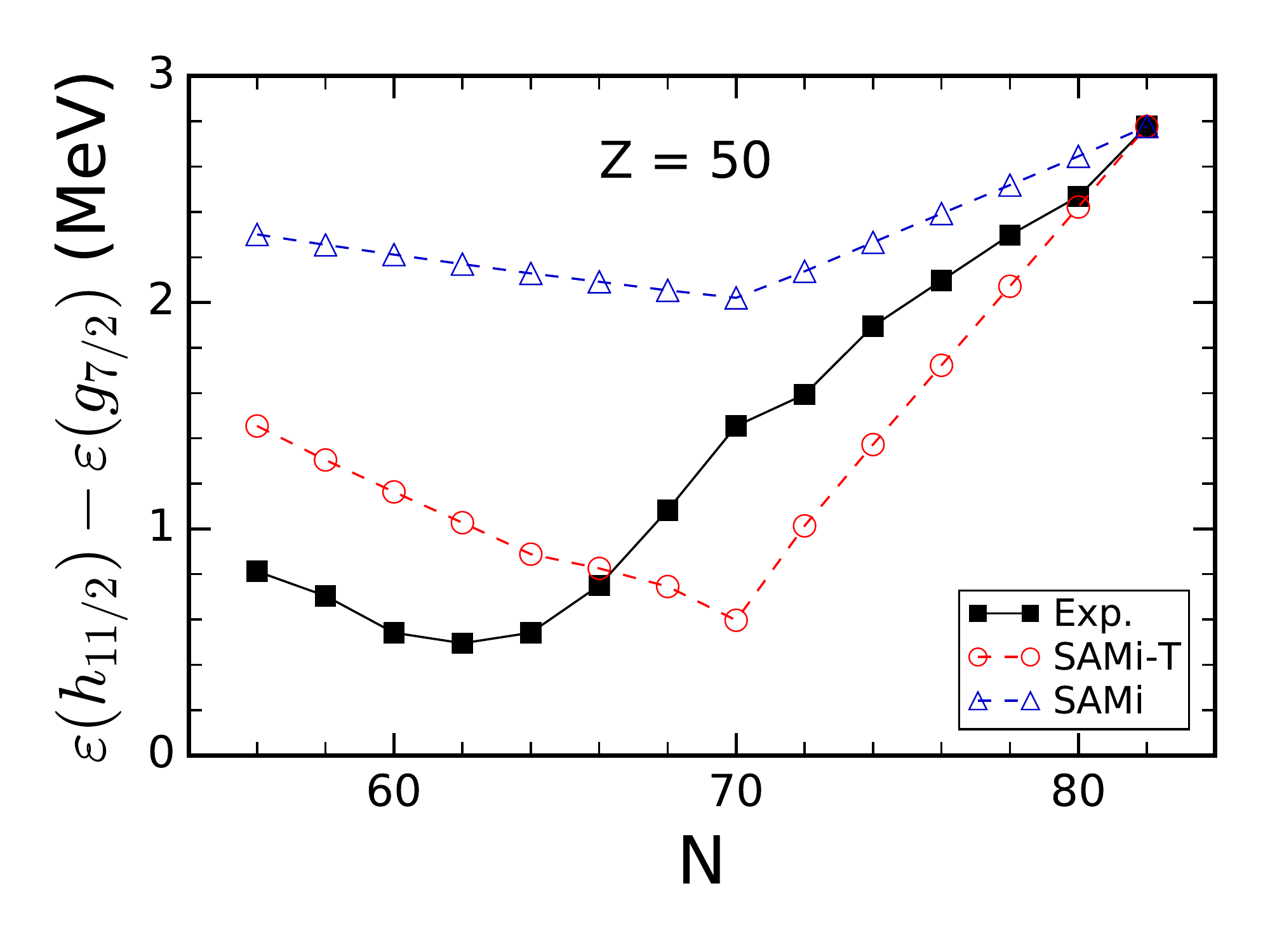}
\caption{Single-particle energy splittings of proton $h_{11/2}$ and $g_{7/2}$ along the $Z = 50$ isotopic chain calculated by SAMi-T and SAMi functionals, in comparison with experimental data \cite{Schiffer2004}.
Theoretical results have been shifted so that the splittings coincide with the data at $^{132}$Sn, see the text for details.}
\label{fig-1}       
\end{figure}

In Figure \ref{fig-1} we show the single-particle energy splittings of proton $h_{11/2}$ and $g_{7/2}$ along the $Z = 50$ isotopic chain calculated by SAMi-T and SAMi functionals, in comparison with experimental data \cite{Schiffer2004}.
In order to focus on the effect of tensor force, which influences the relative change of spin-orbit splittings along the isotopic (or isotonic) chains, the theoretical results have been shifted so that the splittings coincide with the experimental data at $^{132}$Sn.
The shift for SAMi-T is $\delta_{\rm SAMi-T} = 6.8 - 2.8 = 4$ MeV; and for SAMi is $\delta_{\rm SAMi} = 6.5 - 2.8 = 3.7$ MeV.

It can be seen that by including the tensor terms, the relative change of the energy splittings given by SAMi-T is larger from $N = 56$ to $N = 70$, and then to $N = 82$, comparing with SAMi.
This also agrees better with the experimental data than SAMi.
The energy splitting of $h_{11/2}$ and $g_{7/2}$ is determined much by the spin-orbit splittings of the spin doublets $\{h_{11/2},h_{9/2}\}$ and $\{g_{9/2},g_{7/2}\}$.
When the s.o. splittings of these two doublets decrease, the splitting of $h_{11/2}$ and $g_{7/2}$ becomes larger and vice versa.

In the case of Fig.~\ref{fig-1}, the proton numbers are all the same as $Z = 50$ and the neutron numbers change from $N = 56$ to $N = 82$.
Therefore the proton s.o. splittings are much affected by the neutron-proton tensor term, which is determined by the $\beta$ parameter in Eq.~(\ref{eq:alpha-beta}) and the values for the two functionals are \cite{Shen2019}: $\beta_{\rm SAMi-T} = 102$ MeV fm$^5$, $\beta_{\rm SAMi} = 32$ MeV fm$^5$.
In other words, SAMi-T has a much stronger neutron-proton (n-p) tensor effect than SAMi.
For $^{106}$Sn with $N = 56$, the single-particle levels $1g_{9/2}$ and $2d_{5/2}$ are fully occupied while their spin partners $1g_{7/2}$ and $2d_{3/2}$ are empty, thus it is a spin-unsaturated system.
Similarly, $^{132}$Sn with $N = 82$ is also spin-unsaturated as $1h_{11/2}$ is occupied and $1h_{9/2}$ is not.
$^{120}$Sn, with $N = 70$, is spin-saturated as $1g_{7/2}$ and $2d_{3/2}$ are now occupied while $1h_{11/2}$ is empty.
As a consequence, SAMi-T with stronger n-p tensor effect gives relatively smaller proton s.o. splittings for the neutron spin-unsaturated $^{106}$Sn and $^{132}$Sn, and relatively larger splittings for the spin-saturated $^{120}$Sn.
This leads to a relatively larger splitting between $h_{11/2}$ and $g_{7/2}$ at $^{106}$Sn and $^{132}$Sn and relatively smaller splitting as $^{120}$Sn.

\begin{figure}[h]
\centering
\includegraphics[width=7cm]{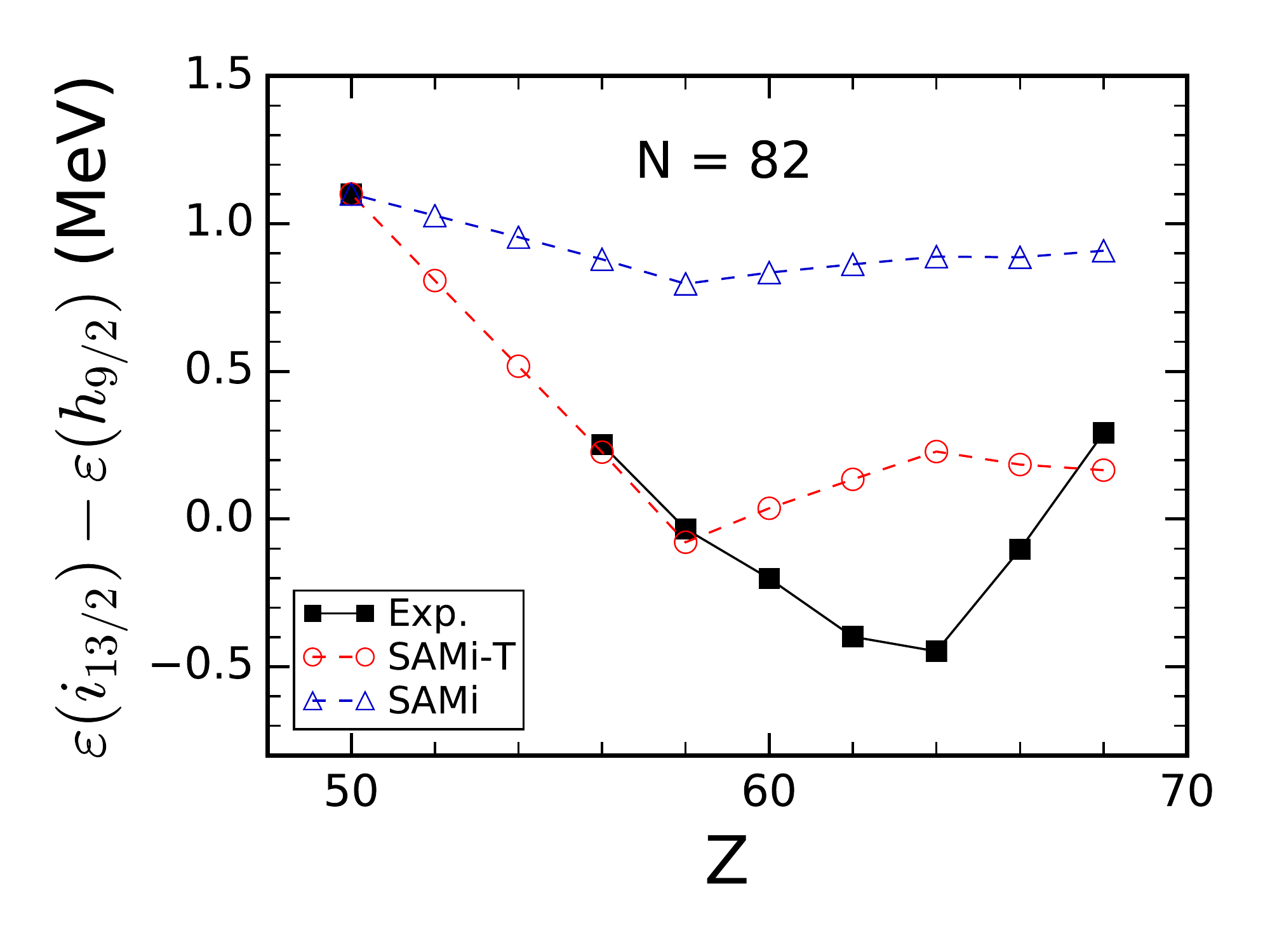}
\caption{Energy splittings of neutron $i_{13/2}$ and $h_{9/2}$ along the $N = 82$ isotonic chain calculated by SAMi-T and SAMi functionals, in comparison with experimental data \cite{Schiffer2004}.
Theoretical results have been shifted so that the splittings coincide with the data at $^{132}$Sn, see the text for details.}
\label{fig-2}       
\end{figure}

A similar analysis can be applied to the mirror case.
In Fig.~\ref{fig-2}, we show the energy splittings of neutron $i_{13/2}$ and $h_{9/2}$ along the $N = 82$ isotonic chain calculated by SAMi-T and SAMi functionals, in comparison with experimental data \cite{Schiffer2004}.
Here the shift for SAMi-T is $\delta_{\rm SAMi-T} = 4.7 - 1.1 = 3.6$ MeV; and for SAMi is $\delta_{\rm SAMi} = 5.3 - 1.1 = 4.2$ MeV.
With the tensor term guided by RBHF calculation, SAMi-T improves much the description of the evolution of s.o. splittings.

In both cases, there is still room to improve.
For example, the minimum of the energy splittings given by the experimental data in Fig.~\ref{fig-1} appears at $N = 62$ and in Fig.~\ref{fig-2} appears at $Z = 64$.
On the other hand, the corresponding minima given by SAMi-T (or SAMi) appear at $N = 70$ and $Z = 58$, respectively.
In the single-particle pictures given by HF calculation, from $N = 52$ to $N = 56$ in the $Z = 50$ isotopes, the $2d_{5/2}$ orbit is being filled.
From $N = 58$ to $N = 64$, the $1g_{7/2}$ orbit is being filled, which is a $j = l-1/2$ orbit and because of the tensor effect discussed above, this decreases the $\{h_{11/2},g_{7/2}\}$ energy splitting.
Furthermore, the occupation of $3s_{1/2}$ and $2d_{3/2}$ orbits will continue to decrease this energy splitting.
In the end, the minimum occurs at $N = 70$.
For $N = 82$ isotones in Fig.~\ref{fig-2}, the situation is slightly different as now the $1g_{7/2}$ orbit is lower than $2d_{5/2}$.
Therefore, from $Z = 52$ on the $1g_{7/2}$ orbit will be filled before $2d_{5/2}$.
Due to the tensor force, these two orbits give contrary effects on the energy splitting of $\{i_{13/2},h_{9/2}\}$ and a minimum occurs at $Z = 58$.
Other features of the energy density functional besides the tensor force can also affect the energy splittings.
To have better agreement with the experiment, further investigations are needed.

\section{Summary}\label{sec-3}

A new Skyrme functional SAMi-T was developed with tensor term guided by \textit{ab initio} relativistic Brueckner-Hartree-Fock calculations.
By fitting to the spin-orbit splittings of neutron-proton drops calculated by RBHF, the tensor terms of SAMi-T are well constrained.
Ground state properties and various giant resonances of finite nuclei can be well described, and tensor term improves the description in some cases \cite{Shen2019}.

The evolution of single-particle energy splittings in the $Z = 50$ isotopic chain and $N=82$ isotonic chain have been studied with SAMi-T functional.
The agreement with experimental data has been improved due to the tensor terms.
%
%
%



\end{document}